\documentclass[twocolumn,floatfix,nofootinbib,british,balancelastpage]{revtex4-1}
\usepackage{graphicx}
\usepackage{rotate}
\usepackage{rotating}
\usepackage{relsize}
\usepackage{lineno}
\usepackage{slashed}
\usepackage{balance}
\usepackage{url}
\usepackage{amsmath,mathrsfs}
\usepackage{amssymb}
\usepackage{longtable}
\usepackage{multirow}
\usepackage{tabularx}
\usepackage{units}
\usepackage[usenames,dvipsnames]{xcolor}
\usepackage{subfigure}
\usepackage{hepunits}
\usepackage{hepnicenames}

\renewcommand{\GeV}{\text{Ge}\hspace{-0.05cm}\text{V}}
\renewcommand{\TeV}{\text{Te}\hspace{-0.05cm}\text{V}}

\newcommand{\qbar}{\bar{q}}
\newcommand{\chibar}{\bar{\chi}}

\usepackage{soul}

\begin{document}

%\titlespacing{\subsection}{0pt}{*0}{*0}

\title{Contact Interactions Probe Effective Dark Matter Models at the LHC }%

\author{Herbi Dreiner }%
\email{dreiner@th.physik.uni-bonn.de}
\affiliation{Physikalisches
  Institut and Bethe
  Center for Theoretical Physics, University of Bonn, Bonn, Germany}

\author{Daniel Schmeier }%
\email{daschm@th.physik.uni-bonn.de}
\affiliation{Physikalisches
  Institut and Bethe
  Center for Theoretical Physics, University of Bonn, Bonn, Germany}

\author{Jamie Tattersall }%
\email{jamie@th.physik.uni-bonn.de}
\affiliation{Physikalisches
  Institut and Bethe
  Center for Theoretical Physics, University of Bonn, Bonn, Germany}

\begin{abstract}
Effective field theories provide a simple framework for probing possible dark matter
(DM) models by re-parametrising full interactions into a reduced number of
operators with smaller dimensionality in parameter space. In many cases these models
have four particle vertices, e.\/g.\/ $q \qbar \chi \chibar$, leading to the pair production of dark
matter particles, $\chi$, at a hadron collider from initial state quarks, $q$.
In this analysis we show that for many fundamental DM models with s--channel DM couplings to
$q \bar{q}$ pairs, these effective vertices must also produce quark contact
interactions (CI) of the form $q \qbar q \qbar$. The respective effective
couplings are related by the common underlying theory which
allows one to translate the upper limits from one coupling to the other. 
We show that at the \textsc{Lhc}, the experimental limits on quark contact 
interactions give stronger translated limits on the DM coupling than the
experimental searches for dark matter pair production.
\end{abstract}


\date{March 15, 2013}%
\maketitle
%\tableofcontents

\section{Introduction}
An explanation of the cosmological dark matter puzzle is one of the most important questions 
facing modern physics. Currently, a compelling explanation of this phenomenon is via 
so-called Weakly Interacting Massive Particles (\textsc{Wimp}s) \cite{Lee:1977ua,Bertone:2004pz,Goldberg:1983nd}. The 
heightened interest in \textsc{Wimp}s is due to the fact that a neutral 
particle which interacts with roughly the strength of the weak force gives the correct DM abundance.

Many experiments are currently under way in the hope of finding 
the DM particle. The most well known are direct detection searches that aim to see DM 
collisions with atomic nuclei \cite{Goodman:1984dc}. Another possible way to see a 
signal could come from the annihilation of DM particles in high density regions of 
the universe, see e.g.~\cite{Bouquet:1989sr}.

More recently, interest has turned to the possibility that particle colliders
could produce dark matter. Under the \textsc{Wimp} hypothesis,
dark matter is assumed to be a neutral stable particle with a mass around the
electroweak symmetry breaking scale. Therefore it is natural to ask if the
signatures of these particles could be seen at \TeV{} scale colliders. Using a
model independent or effective theory approach
\cite{Kurylov:2003ra,Beltran:2008xg,Agrawal:2010fh}, it is possible to
relate expected production cross-sections to the relic density
\cite{Birkedal:2004xn} and/or to the direct detection cross-section
\cite{Cao:2009uw,Beltran:2010ww,Bai:2010hh}. Since the dark matter candidates
are only weakly interacting, observations require Standard Model particles to
be produced in association. These are usually a photon or jet that has been
radiated from the initial state and lead to a mono-photon (or -jet) topology. Many studies have 
now investigated these kinds of signals in a model independent or effective theory 
approach \cite{Bartels:2007cv,Konar:2009ae,Cao:2009uw,Bai:2010hh,Fox:2011pm,Goodman:2010ku,
Goodman:2010yf,Beltran:2010ww,Rajaraman:2011wf,Bai:2012he,Cheung:2012gi,Haisch:2012kf,MarchRussell:2012hi,
Mambrini:2011pw,Fox:2012ru,Frandsen:2012rk,Tsai_New,Dreiner:2012xm,Chae:2012bq}. In the case of 
Supersymmetry the same signal can be used to search for compressed 
spectra \cite{Belanger:2012mk,Dreiner:2012gx,Dreiner:2012sh}. 

In the effective field theory approach, higher dimensional operators are employed as an approximation 
to a full theory that includes dark matter. For this approach to be valid, the full theory must 
contain at least one additional heavy particle that mediates the interaction between the Standard 
Model and dark matter. Consequently, it is possible that the presence of these new states may 
have observable effects at colliders and these should be explored. For example, if the interaction 
between dark matter and the Standard Model is via an s-channel process, the mediator 
could produce a di-jet resonance \cite{Frandsen:2012rk,Tsai_New}. In this case, the search for di-jet 
resonances is often more powerful than the more direct mono-jet searches for the dark matter particle.
However, they are limited to models where the mediator can be produced on-shell at the collider in 
question. In this case, the validity of using an \emph{effective theory} to 
describe the dark matter production 
must be called in question.  

Instead of using a di-jet resonance search to constrain the effective model, we propose to 
look for deviations from the Standard Model in $q\bar{q}$ contact interactions. In the case of an 
s-channel mediator, the operator will lead to a deviation from standard \textsc{Qcd} 
interactions and thus can be searched for. In fact \textsc{Atlas} and 
\textsc{Cms} \cite{ATLASContactDijets,ATLASContactMuons,CMSContactDijets,CMSContactMuons} have 
searches of this kind in various final states.

In this study we focus on deviations in the high energy di-jet
spectrum. We choose di-jets, since if a mediator couples to a $q\bar{q}$
initial state, it is guaranteed to also mediate the production of di-jets. We
can compare to the mono-jet analyses by noting that the same interaction
between the mediator and initial state quarks must also exist
there. Consequently, if we assume that the interaction between the mediator
and the dark matter particles is perturbative $(g\leq\sqrt{4\pi})$, the
mono-jet search sets a limit on the interaction between the mediator and
quarks. We show that in the region of parameter space where the effective
theory is valid (mediator mass $\gtrsim$ \unit{1}{\TeV}) and perturbativity is not violated, the contact interaction
search leads to the most stringent limit.  

We first give an example fundamental model and show how an
effective theory of interactions can be derived in the limit of a large
mediator mass in Sec~\ref{sec:modelintro}. Next we explain the limits on this
effective theory coming from both contact interactions and mono-jet searches
for dark matter in Sec.~\ref{sec:EffLimits}. We continue in
Sec.~\ref{sec:ExpLimits} with the comparision of limits on the effective couplings and show 
that at the \textsc{Lhc} contact interaction bounds lead to more stringent limits. Different fundamental theories
may be expected to have different bounds on the underlying couplings and we address these questions 
in Sec.~\ref{sec:leptoncoupling}. We conclude the paper in Sec.~\ref{sec:conclusion}.

\section{Effective Couplings From a Fundamental Model}
\label{sec:modelintro}
We start with a simple formulation of an example model to describe the interaction of 
a new dark matter particle $\chi$ with Standard Model quarks $q$. We choose $\chi$ to 
be a Dirac fermion and analyze pair production $q q \rightarrow \chi \chi$ from 
initial state quarks, via a heavy vector mediator $V$ from an U(1) gauge theory. A particle $X$ is assumed 
to have mass $M_X$. We consider the following Lagrangian for this model,
\begin{align}
\mathscr{L}_\text{UV} &= \bar{q} ( i \slashed{\partial} - M_q ) q + \bar{\chi} ( i \slashed{\partial} - M_\chi ) \chi  \nonumber \\
 &+ \frac{1}{2} M_V^2 V_\mu V^\mu - \frac{1}{4} V^{\mu \nu} V_{\mu \nu}  \nonumber \\
 &- g_q \bar{q} \gamma^\mu P_L q V_\mu -  g_\chi \bar{\chi} \gamma^\mu P_L \chi V_\mu , \label{eqn:modellagrangian}
\end{align}
where we have used the projection operator  
\begin{equation}
  P_L \equiv \frac{(1-\gamma^5)}{2}.
\end{equation}
The first four terms include both kinematic and mass terms for all the fields 
(with the standard Abelian field strength tensor $V^{\mu \nu} \equiv \partial^\mu V^\nu - \partial^\nu V^\mu$ for 
the vector mediator). The last terms describe chiral interactions of the vector particle $V^\mu$ with 
both fermions $\chi$ and $q$ via dimensionless coupling strengths $g_q$ and $g_\chi$. The 
particular choice of a chiral interaction leads to effective operators that are commonly
analysed in experimental studies, e.\/g.\/ \cite{ATLASContactDijets, CMSContactDijets}. We 
consider different operators in section \ref{sec:leptoncoupling}.

 The DM particle 
$\chi$ is assumed to interact with the Standard Model only by exchanging the new mediator $V$, i.\/e.\ it 
is uncharged under any Standard Model gauge group and  neither couples to the respective gauge bosons nor the Higgs particle. 

\begin{figure}
\centering         
\subfigure[\ Elastic quark scattering (plus a corresponding t-channel contribution).]{ \includegraphics[width=0.45\columnwidth]{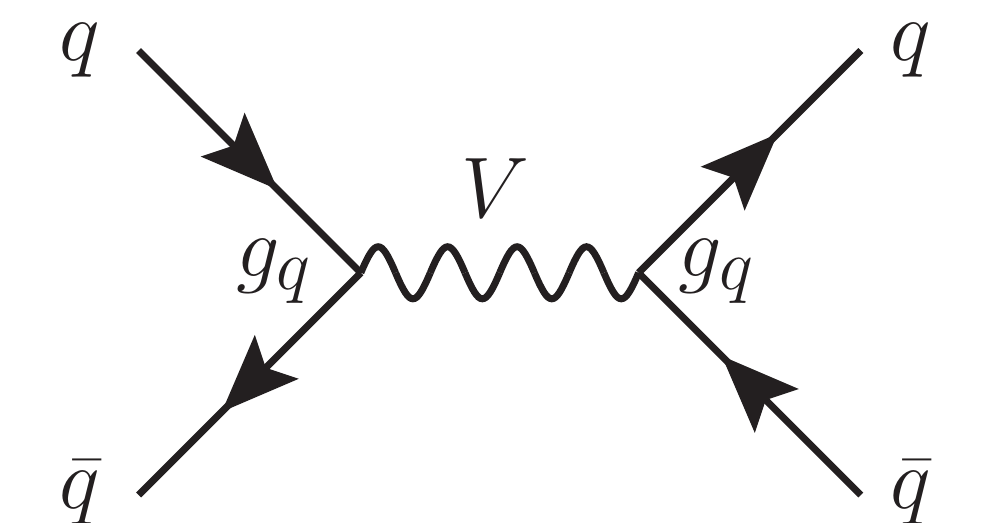}}
\subfigure[\ Pair production of $\chi$.]{ \includegraphics[width=0.45\columnwidth]{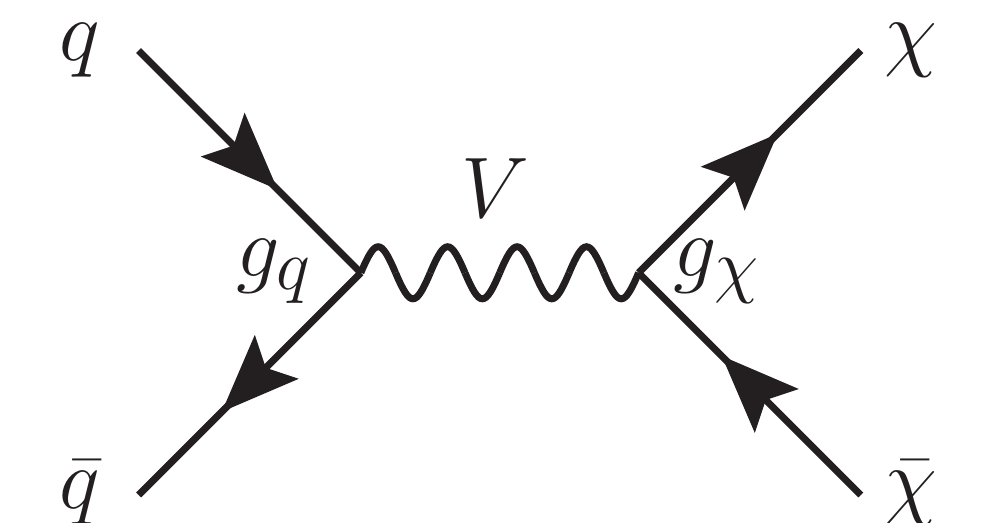}}
\caption{New interaction modes for quarks in the initial state, given by the model introduced in (\ref{eqn:efflagrangian}).}
\label{img:interactions}
\end{figure}

The new mediator leads to new interaction channels for the Standard Model quarks, which are shown 
in Fig.~\ref{img:interactions}. At a hadron collider, an off-shell mediator that is created by two 
initial state quarks can either produce a pair of quarks, describing elastic quark scattering, 
or produce a pair of the new particle $\chi$. Since both processes depend on the 
strength of the initial state coupling  $g_q$, their cross sections are related. 

If we now assume that the mass of the mediator, $M_V$, lies far beyond the accessible center of mass energy $\sqrt{\hat{s}}$  
of the partons in any scattering process we want to analyse at a hadron collider, we can integrate out the vector field and expand the remainder of the effective Lagrangian up to leading order in $\hat{s}/M_V^2$ (see e.\/g.\ \cite{Haba:2011vi}),
\begin{align}
\mathscr{L}_\text{eff} &= \bar{q} ( i \slashed{\partial} - M_q ) q + \bar{\chi} ( i \slashed{\partial} - M_\chi ) \chi \nonumber \\
&- \frac{g_q^2}{2 M_V^2} \bar{q}_L \gamma^\mu q_L \bar{q}_L \gamma_\mu q_L - \frac{g_q g_\chi}{M_V^2} \bar{q}_L \gamma^\mu q_L \bar{\chi}_L \gamma_\mu \chi_L  \nonumber \\
& - \frac{g_\chi^2}{2 M_V^2} \bar{\chi}_L \gamma^\mu \chi_L \bar{\chi}_L \gamma_\mu \chi_L, \label{eqn:efflagrangian}
\end{align}
with the left--handed component of the quark field $q_L \equiv P_L q$.
The last term describes the scattering of the dark matter particle $\chi$ with itself, which is 
of no interest in this analysis and is therefore omitted henceforth. We combine the pre-factors 
of the two remaining effective vertices by defining the effective couplings $G_q \equiv g_q^2/M_V^2$, 
describing a contact interaction (CI) between four Standard Model quarks, and $G_\chi \equiv g_q g_\chi / M_V^2$, 
which gives the scattering strength between quarks and the DM particle $\chi$. 

To be consistent with the perturbative approach of using tree-level diagrams only, the 
dimensionless couplings $g$ must not be larger than $\sqrt{4 \pi}$. Thus, in addition to the 
restriction $M_V^2 \geq \hat{s}$ demanded for the effective approximation to be valid, 
only the limited parameter space $0 < G_i < 4 \pi/ \hat{s}$ is allowed for both effective couplings $G^i$.

\newcommand{\Geff}{G}
\section{Experimental Limits on the Effective Couplings}
\label{sec:EffLimits}
The two effective couplings we derived have to be probed differently at a hadron collider. Firstly, $\Geff_q$ 
describes the elastic scattering of quarks and can be analysed by looking for deviations compared to Standard Model 
predictions for high energy di-jet production. This analysis has been performed by both 
the \textsc{Atlas} \cite{ATLASContactDijets} 
and \textsc{Cms} \cite{CMSContactDijets} collaborations at the \textsc{Lhc}. Since there also exist Standard Model diagrams 
for this type of scattering, limits on $\Geff_q$ depend on how the Standard Model terms interfere with the new contribution 
of the effective operator. We conservatively take the lowest limits given for destructive interference, 
which \textsc{Cms} quotes as,
\begin{align}
\Geff_q \leq 4 \pi (\unit{7.5}{\TeV})^{-2} \label{eqn:limitOnGpsi}
\end{align}
at 95\% CL, determined with an integrated luminosity of \unit{2.2}{\femto\barn}$^{-1}$ at \unit{7}{\TeV} center of mass energy.

On the other hand, $\Geff_\chi$ describes dark matter pair production. These particles are usually invisible at the LHC, 
since they do not interact significantly with the detector at the \textsc{Lhc} due to their small coupling to the Standard Model. 
Mono-photons \cite{CMSDarkmatterPhotons, ATLASDarkmatterPhotons} or mono-jets \cite{CMSDarkmatterJets,ATLAS:2012ky} 
(radiated from the initial state) are characteristic for this kind of interaction and have been probed by both 
experiments\footnote{To be precise, the limits have been determined for the 
vertices $\qbar \gamma^\mu q \chibar \gamma_\mu \chi$ and $\qbar \gamma^\mu \gamma^5 q \chibar \gamma_\mu \gamma^5 \chi$ 
individually. 
However, for a large mass range the bounds are similar. Therefore we assume the same limits on the 
coupling $\bar{q}_L \gamma^\mu q_L \bar{\chi}_L \gamma_\mu \chi_L$.}. The currently 
strongest upper bound on  $\Geff_\chi$ is given by a mono-jet analysis of \textsc{Cms}
\cite{CMSDarkmatterJets} for an integrated luminosity of \unit{5.0}{\femto\barn}$^{-1}$ at \unit{7}{\TeV},
\begin{align}
\Geff_\chi \leq (\unit{765}{\GeV})^{-2} \label{eqn:limitOnGchi},
\end{align} which holds for $M_\chi = \unit{10}{\GeV}$ at 90\% CL. Different (larger or smaller) values for $M_\chi$ lead to weaker bounds. 
%Note that due to the smaller level of confidence, this limit is less stringent than the CI bounds in (\ref{eqn:limitOnGpsi}).

\section{Comparing Experimental Limits}
\label{sec:ExpLimits}
The quoted limits on $\Geff_q$ and $\Geff_\chi$ differ significantly, due to the very different techniques 
involved in the respective analyses. However, the two effective couplings have common ingredients which 
implicitly relate them. Consequently, we may reasonably translate the limit from $\Geff_q$ into an upper bound 
on $\Geff_\chi$ and see how this bound compares to the experimental limit given in (\ref{eqn:limitOnGchi}).  

Since $\Geff_\chi$ depends on $g_\chi$ whereas $\Geff_q$ does not, there is no 1:1-correspondence 
between the two effective couplings and they are \textit{a priori} independent. However, we only 
have restricted parameter values for the coupling constants $g_\chi$ and $g_q$ to be in agreement with the 
perturbative picture. Taking the definition for $\Geff_\chi$ and restricting $g_\chi, g_q \leq \sqrt{4 \pi}$ by 
perturbation theory, it follows that,
\begin{align}
\Geff_\chi \leq \frac{4 \pi}{M_V^2}. \label{eqn:gchiperturbativelimit}
%\Geff_\chi \leq 4 \pi M_V^{-2}. \label{eqn:gchiperturbativelimit}
\end{align}
Furthermore we may relate $\Geff_\chi$ to $\Geff_q$ in order to apply the experimental limit known for $\Geff_q$. 
According to the definitions of the two effective couplings, it follows that,
\begin{align}
\Geff_\chi = \frac{g_\chi}{M_V} \sqrt{\Geff_q}. \label{eqn:geffrelation}
%\Geff_\chi = g_\chi M_V^{-1} \sqrt{\Geff_q}. \label{eqn:geffrelation}
\end{align}
With the experimental limits on $\Geff_q$ given in  (\ref{eqn:limitOnGpsi}) and 
the perturbative restriction $g_\chi \leq \sqrt{4 \pi}$, we find,
\begin{align}
\Geff_\chi \leq \frac{1}{M_V} \frac{4 \pi}{\unit{7.5}{\TeV}}. \label{eqn:gchigpsilimit}
%\Geff_\chi \leq 4 \pi (M_V \cdot \unit{7.5}{\TeV})^{-1}. \label{eqn:gchigpsilimit}
\end{align}
\begin{figure}
\centering
\includegraphics[width=0.99\columnwidth]{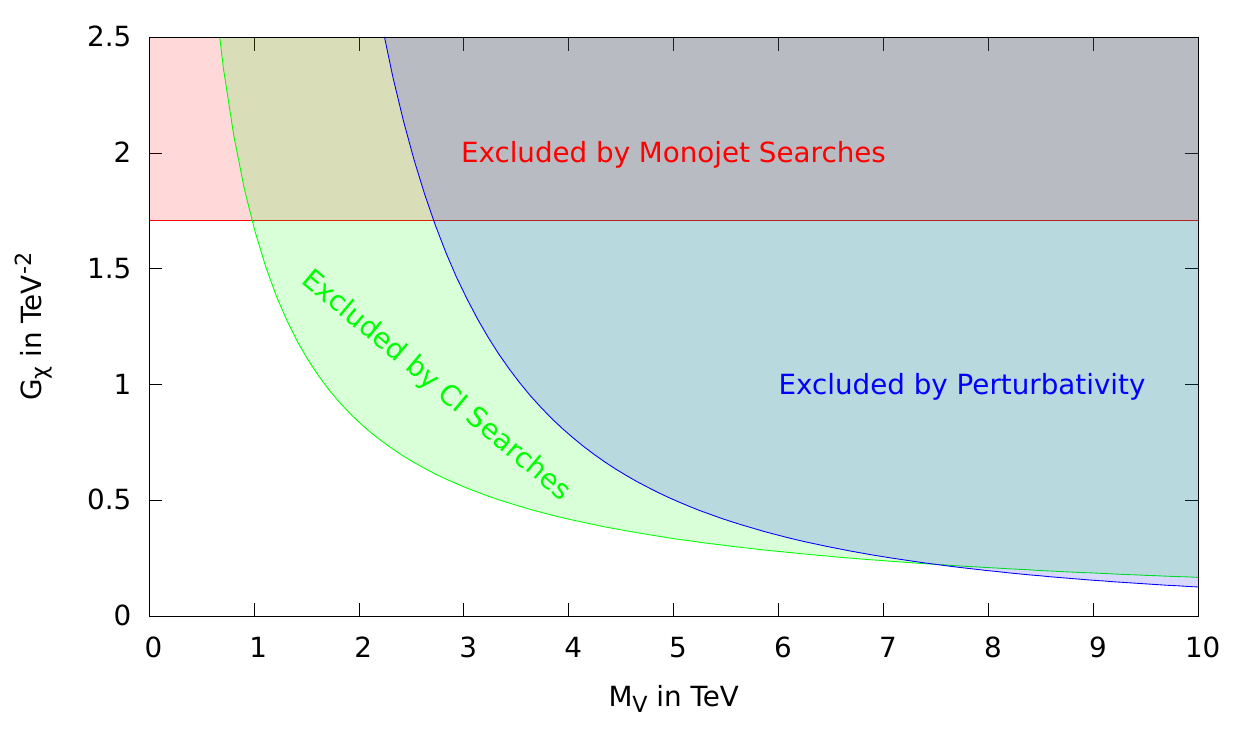}
\vspace{-0.5cm}
\caption{Exclusion limits on the effective coupling constant $\Geff_\chi$ for given mediator mass $M_V$ 
according to experimental limits from mono-jet searches (red), experimental limits on the contact 
interaction $\Geff_q$ (green) and the perturbative restriction on the fundamental coupling constants $g_i$ 
(blue). The bound from monojet searches assumes a dark matter mass $M_\chi =
\unit{10}{\GeV}$, which is the most optimistic scenario and leads to the
strongest bound. The upper limit from contact interactions assumes destructive
interference which gives the most conservative limit. For mediator masses consistent with the effective approach ($M_V \gtrsim \unit{1}{\TeV}$), we see the bound from 
contact interactions is most stringent.}
\label{fig:exclusionlimits_1}
\end{figure}
In Fig.~\ref{fig:exclusionlimits_1} we compare the excluded
parameter regions in the $\Geff_\chi$--$M_V$ plane according to the different
restrictions (\ref{eqn:gchiperturbativelimit})--(\ref{eqn:gchigpsilimit}). It can be seen 
that in the mediator mass range from \unit{1}{\TeV} up to \unit{7}{\TeV}, the translated 
experimental limit on the quark contact interaction $\Geff_q$ gives the strongest restrictions 
on the parameter space of the effective theory for the dark matter particle $\chi$. In particular, 
the limits are stronger than the experimental constraints on $\Geff_\chi$ from mono-jet searches. 

For larger mediator masses, demanding that the theory is perturbative gives stronger upper limits than both of the 
experimental searches. However, the perturbative upper bound is static whereas the experimental sensitivity will gradually 
improve over time as more data is collected.

In the small $M_V$ limit below \unit{1}{\TeV}, experimental limits on $\Geff_q$ can only give 
weak statements on the allowed parameter space and the mono-jet searches give the strongest 
exclusion limit. Unfortunately, the typical energies involved in general scattering 
processes at the \textsc{Lhc} are likely to be at or above the \mbox{\TeV-scale}. According to the 
requirements of an effective theory to be valid, mediator masses below \unit{1}{\TeV} cannot be 
analysed reasonably in that framework and experimental limits cannot be trusted anymore. However, 
we note that the searches for di-jet \textit{resonances} can lead to bounds in these mass 
ranges and in many cases, these are more stringent than those coming from mono-jet searches 
\cite{Frandsen:2012rk,Tsai_New}. 

We thus conclude that for mediator masses that allow the application of an 
effective approximation, experimental limits from contact interaction searches or perturbativity 
bounds put stronger restrictions on the allowed parameter space of an effective dark matter theory 
than mono-jet searches. 
\section{Applicability to Other Effective Models}
\label{sec:leptoncoupling}
In the preceding analysis we considered the specific effective model given in (\ref{eqn:modellagrangian})
as an example. We now discuss how a different model might change the statements given so far.
\subsection{Different Spins for $\chi$}
If the introduced new particle $\chi$ is not fermionic but a scalar or vector particle, the 
Lorentz structure of its coupling to the mediator changes (see e.\/g.\/ \cite{Dreiner:2012xm}). However, this neither 
affects the effective Standard Model contact interaction nor the definitions of the effective 
coupling constants $\Geff$. Therefore both limits (\ref{eqn:gchiperturbativelimit}) and (\ref{eqn:gchigpsilimit}) remain. 

For the mono-jet/mono-photon searches, a different spin for $\chi$ might change the total 
orbital angular momentum of the final state and therefore affect the  kinematics of the events. 
This would lead to a change in the expected signal and background in the kinematic 
region of interest such that the derived limits on $\Geff_\chi$ could change. However, the 
cross section for effective theories has its maximum for small angles and energies of the radiated object. In that 
kinematic regime, radiated objects are mainly described by spin-independent splitting functions, 
i.\/e.\/ the Weizs{\"a}cker-Williams distribution in case of soft photons \cite{Weizsacker:1934sx, PhysRev.45.729} or the solution of the 
\textsc{Dglap}-equations for soft gluons \cite{Altarelli:1977zs, Dokshitzer:1977sg, Gribov:1972ri}.  Thus we do not expect a significant impact of the spin 
structure on the interaction and therefore no significant effect on the experimental limits.
\subsection{Different Spins for the Mediator}
A different spin for the mediator changes both effective operators and therefore affects the 
 signal for both experimental searches. For the mono-jet/mono-photon searches we still 
do not expect large differences for the same reason we gave in the case of a different spin 
for $\chi$. The di-jet analyses usually examine the angular distribution of the two final state 
objects, which changes with the spin of the mediator. However, the expected signal 
distributions should still be distinguishable from the \textsc{Qcd} background\footnotemark, 
such that we expect the derived limits to not change drastically. 
\subsection{Changing the Interaction Vertices}
Changing any vertex from a chiral interaction to e.\/g.\/ vector or axial-vector like 
couplings usually affects the dependence of any related cross section on the initial 
particle spin polarisation. However, since the \textsc{Lhc} only measures unpolarised 
cross sections, these changes are not expected to change the respective limits on the 
effective couplings and thus keep the aforementioned statements valid.

If either the mediator or the dark matter particle is a fermion, a t-channel 
interaction of the form $V^\mu(\qbar \gamma_\mu \chi + \chibar \gamma_\mu q)$ is also possible. 
In that case, the effective approximation does not give any $q^4$ interaction to leading order in $\hat{s}/M_V^2$ such that CI measurements are not sensitive anymore. Limits from mono-jet searches 
then give the strongest restrictions in parameter space in the regions allowed by perturbation theory.

We note that if the mediator $V$ and the DM candidate $\chi$ are either both scalars 
or vectors, their respective coupling $\chi \chi V$ has mass dimension 3 such that the 
corresponding coupling constant $g$ is not dimensionless anymore. In that case neither 
the perturbative upper limit of $\sqrt{4 \pi}$ can be applied, nor do we expect the 
experimental limits to stay constant since the cross sections formul\ae{} most likely 
change non-trivially due to the different mass dimension of $\Geff$. 
\footnotetext[\value{footnote}]{Both di-jet analyses \cite{CMSContactDijets, ATLASContactDijets} 
look at the distribution of $\chi \equiv \exp (|y_1-y_2|)$, where $y \equiv 1/2 \ln ( (E+p_z)/(E-p_z) )$ 
is determined from the two highest $p_T$ jets. It behaves quite uniformly for the dominating \textsc{Qcd} 
background, whereas it strongly peaks to small values in case of various effective s-channel contact interactions.}

\subsection{Coupling to Standard Model Leptons}
\begin{figure}
\centering
\includegraphics[width=0.99\columnwidth]{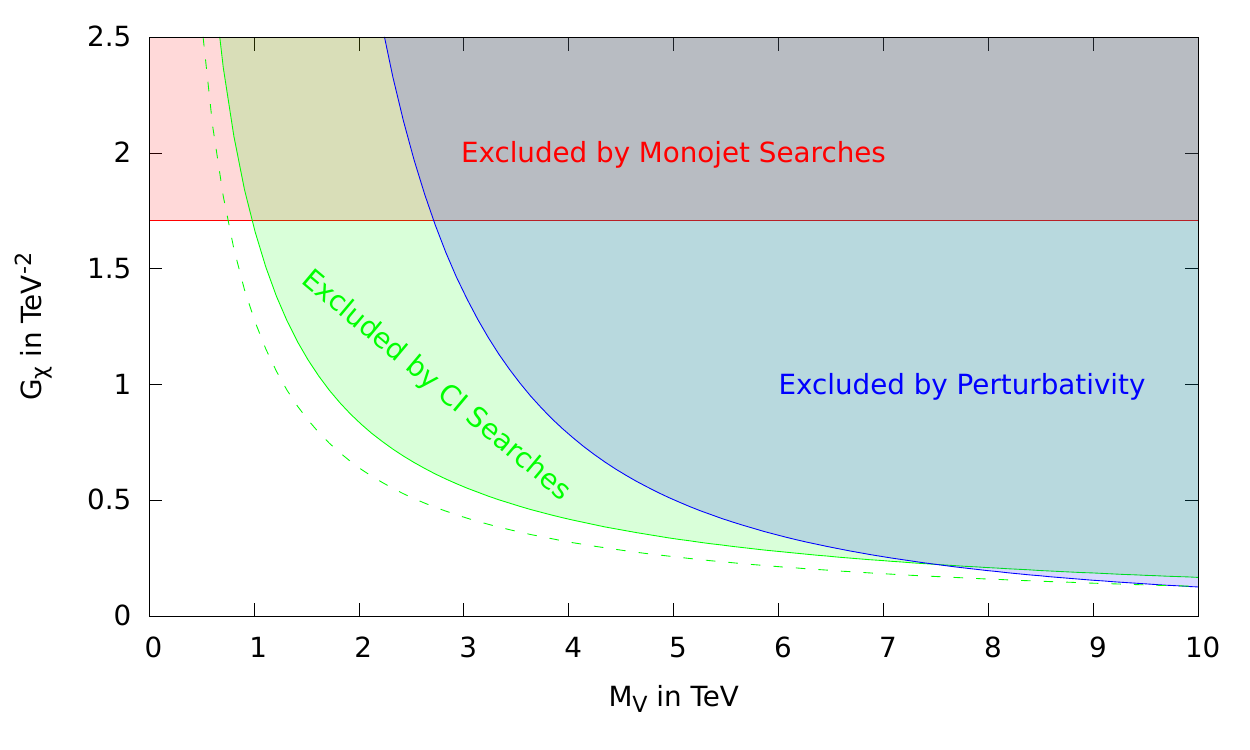}
\vspace{-0.5cm}
\caption{Exclusion limits on the effective coupling constant $\Geff_\chi$ for given mediator mass $M_V$ 
as in Fig.~\ref{fig:exclusionlimits_1}. The dashed line shows the improved CI limits from dilepton 
searches in case of a universal coupling of the mediator to quarks and leptons.}
\label{fig:exclusionlimits}
\end{figure}

The new mediator may also have an additional coupling to the leptonic sector of the Standard Model, from 
which one would expect an effective contact interaction of the form $\qbar q \bar{l} l$. If the mediator 
couples universally to all Standard Model particles, this vertex would be described by the same effective 
coupling $\Geff_q$ as the CI quark interaction we analysed before. This new coupling could then be 
probed by di-lepton searches \cite{CMSContactMuons, ATLASContactMuons}. The \textsc{Atlas} collaboration 
quotes a combined limit given by both di-electron and di-muon searches of,
\begin{align}
\Geff_q \leq 4 \pi(\unit{9.8}{\TeV})^{-2} \label{eqn:dileptonlimit}
\end{align}
at 95\% CL, determined with an integrated luminosity of
\unit{5.0}{\femto\barn}$^{-1}$ at \unit{7}{\TeV}. The resulting limits on $\Geff_\chi$ are shown
in Fig.~\ref{fig:exclusionlimits}. They are more restrictive than the di-jet
limits given in (\ref{eqn:limitOnGpsi}), which is not only due to the larger
integrated luminosity but also because of the cleaner final state. However, the di-jet limits are valid 
for a larger set of models, i.\/e.\/ for all that introduce quartic quark self interaction, 
whereas the dilepton limits need to assume quark-lepton universality. 
\subsection{Coupling to Gluons}
Since we are discussing the experimental results of a hadron collider, not only quark couplings 
but also the interaction with gluons can be probed. There are different possibilities to 
construct effective interactions of gluons with the particle $\chi$ (see e.\/g.\/ \cite{Beltran:2010ww}), 
that may lead to anomalous 4-gluon couplings. However, there are some drawbacks 
which make the direct analogy difficult:
\begin{itemize}
\item In our approach, we start with the underlying renormalizable ultraviolet complete theory. Writing down an interaction for gluons with any mediator $V$ should 
both keep the dark matter candidate a gauge singlet and be in agreement with SU(3) color symmetry. This demands the existence of a $\chi \chi V V$ term, which leads to more involved effective theories due to the necessity of 
exchanging mediators in pairs. It can therefore be expected that the relation between the two 
effective coupling constants $\Geff^g$ and $\Geff_\chi$ becomes non-trivial and does not allow for 
an easy comparison as above.
\item Possible measurements of different gauge invariant anomalous gluon self couplings have been 
analysed in \cite{AnomalousGluons, Dreiner1992441}. It is stated that di-jet analyses are practically impossible 
in that framework, since new operators contribute to the order $1/M_V^4$, whereas for quark 
operators as in (\ref{eqn:efflagrangian}) they already arose at order $1/M_V^2$. This leads 
to a very small expected signal which is difficult to probe at \textsc{Lhc} energies. 
Other proposed methods might form better alternatives but their sensitivity is still expected to be small. 
\end{itemize}
We therefore do not expect that our statement about the relative strength of mono-jet and CI 
searches holds in case of gluonic operators.
\section{Conclusion} 
\label{sec:conclusion}
In this study we have compared the bounds placed on effective theories of dark
matter from mono-jet searches and contact interactions. We have shown that for
models which dominantly interact at the \textsc{Lhc} via a $q\bar{q}$ initial
state and an s-channel mediator, the bounds from contact interactions are the
most stringent in the regions of parameter space where the effective theory is
valid and the couplings are perturbative. For these models, the contact interaction 
searches can probe mediator masses up to \unit{7}{\TeV}. For lower mediator masses, 
the limit from contact interactions is the most constraining as long as the mediator mass 
is above $\sim$\unit{1}{\TeV}. Due to the nature of the contact interaction being probed 
there is no dependence of the dark matter mass on the analysis.

Additionally we have also commented on the applicability of these limits to other effective 
models. We believe that the conclusions presented will only depend weakly on the spin of 
the dark matter and/or the mediator. We again note that the limit is only valid for s-channel 
mediators but any other change to the interaction vertex will leave our conclusions unchanged. 
In the case that the mediator couples to leptons in the final state, the bounds from contact 
interactions may even be increased. However, we unfortunately find that for gluonic couplings 
no definitive statements can be made from a contact interaction analysis.
\section*{Acknowledgements}
\noindent We would like to thank Lorenzo Ubaldi for early discussions regarding this work. The work has been supported
by the BMBF grant 00160200 and the \textsc{Dfb Sfb/Tr33} `The
Dark Universe'.

\bibliographystyle{h-physrev5}
\bibliography{contact}

\end{document}